# A new approach for detecting scientific specialties from raw cocitation networks

Matthew L. Wallace and Yves Gingras

Observatoire des sciences et des technologies (OST), Centre interuniversitaire de recherche sur la science et la technologie (CIRST), Université du Québec à Montréal, Case Postale 8888, succ. Centre-Ville, Montréal (Québec), H3C 3P8, Canada. E-mail: mattyliam@gmail.com; gingras.yves@uqam.ca

Russell Duhon

School of Library and Information Science, Indiana University, 1320 East 10th Street, LI 011, Bloomington, IN, 47405-3907, USA. E-mail: rduhon@indiana.edu

## Abstract

We use a technique recently developed by Blondel et al. (2008) in order to detect scientific specialties from author cocitation networks. This algorithm has distinct advantages over most of the previous methods used to obtain cocitation "clusters", since it avoids the use of similarity measures, relies entirely on the topology of the weighted network and can be applied to relatively large networks. Most importantly, it requires no subjective interpretation of the cocitation data or of the communities found. Using two examples, we show that the resulting specialties are the smallest coherent "group" of researchers (within a hierarchy of cluster sizes) and can thus be identified unambiguously. Furthermore, we confirm that these communities are indeed representative of what we know about the structure of a given scientific discipline and that, as specialties, they can be accurately characterized by a few keywords (from the publication titles). We argue that this robust and efficient algorithm is particularly well-suited to cocitation networks, and that the results generated can be of great use to researchers studying various facets of the structure and evolution of science.

## 1. Introduction

There is an increasingly large amount of literature devoted to the treatment of cocitation data, either of papers, authors or journals. Most of these studies use this readily-available information in order to map the structure of science or identify different clusters of scientific research. The idea behind this type of work, initially developed and used by H. Small and others (Small, 1973; Marshakova, 1973; Small & Griffith, 1974; White, 1981; White & McCain, 1981; Small & Sweeney, 1985; Bayer et al., 1990) is to use cocitations as the foundation of a *conceptual* network that evolves in time based on the choices (citation practices) of scientists themselves (Small, 1978). Often, these conceptual domains turn out to be very similar to what have come to be known as invisible colleges of scientists (Crane, 1972; Mullins et al., 1977).

Several recent developments in the visualization of networks, for instance, have begun to tackle the problem of large-scale, complex cocitation networks (Chen, 2004; Small, 2006; Boyack et al., 2007; Klavans & Boyack, 2006). In general, however, the treatment of cocitation data is not straightforward: a recent debate on the appropriate similarity measures to evaluate the "proximity" of agents (Ahlgren et al., 2003, 2004; White, 2003; Bensman, 2004; Leydesdorff & Vaughan, 2006; Leydesdorff, 2008) highlights the need for a more clear-cut, unbiased methodology for detecting local research communities corresponding to scientific specialties. Indeed, the problem of cluster selection remains important not only for cocitation networks, but for network theory in general (Schaffer 2007, Hsieh & Magee, 2008). In this paper, we evaluate a new community detection method (Blondel et al., 2008) used for identifying, *without any free parameters, pre- or post-processing of data*, scientific specialties for any given cocitation network. We have tested the method using raw author cocitation data from a variety of disciplines or from science as a whole, and from various time periods over the past century, all of which reveal the presence of distinct and readily-identifiable scientific specialties. Here we present two test cases of author cocitations, and show what this new method can uncover about the structure of science and how it may be useful not only in the field of bibliometrics, but also in the history and sociology of science.

Our new approach is motivated by four requirements with respect to the clustering of cocitation networks. First, the weight of the links between authors (number of cocitations) is crucial; this is where most of the information is contained. Therefore, any network-based approach must be able to take into account not only the existence of links between authors, but also how strong these links are. Second, there should be no "choice" made by the user regarding which clusters to identify, nor should there be any *a priori* limitations as to the number of communities or as to their population; a single optimization function or algorithm should provide an independent division of the network. Third, aside from the case of extremely large networks (see Section 3, below), there should be no restrictions on the size or topology of the network used. Naturally, some networks have a more clear-cut community structure than others and this should be reflected in the method's output. Finally, there should not be any *a priori* assumptions on the networks themselves. In other words, they need not be altered in any way before applying the algorithm and *only their inherent structure should be used to determine how they should be partitioned*. The algorithm created by Blondel et al. (2008) explicitly satisfies these conditions. In the following section, we recall the main methods used until now and discuss their limits. In Sections 3 and 4, we will present our methods and results, and compare them with those obtained from other techniques which have previously been applied to cocitation networks.

## 2. The advantages and drawbacks of the dominant decomposition methods

It is worthwhile to briefly examine some of the most successful community detection methods, especially those that have been (or could be) applied to cocitation networks. One must keep in mind that conceptual networks are *not* social networks, so care should be taken regarding which techniques may be transferred between the two. Although this is not intended to be an exhaustive review, all existing methods for *weighted* networks contain, to our knowledge, at least one of the pitfalls discussed here. The simplest approaches consist in using either the absolute or relative cocitation counts. While this has proven useful for certain networks (for good reviews on the subject, see Small, 1985; Gmür, 2003), a threshold must be imposed to differentiate the various sub-groups and the divisions are often contingent on the presence of several dominant authors.

Generally, some indirect measure must be used to compare the members of a network. All similarity measures (e.g. Pearson, Jaccard, cosine) also require, by definition, an arbitrary or semi-arbitrary threshold in order to partition a given network. While visual inspection of the similarities or the application of multidimensional scaling (MDS) provides important insight into the relationships between elements of a network, one must generally resort to decomposition via agglomerative hierarchical clustering (AHC) in order to determine the "optimal" partitioning scheme. This is further complicated by the fact that these methods often use indices that are highly similar to one another. Our task here is not to evaluate the relevance of these measures, nor to discuss their fine-tuning (e.g. how to treat diagonal values of the cocitation matrix). Rather, we contend that this *multiplicity* of perspectives and the sensitivity of the network decomposition to a given measure of similarity are, in our opinion, unsatisfactory, especially when it comes to more complex networks, whose subgroups cannot be confirmed intuitively or by visual inspection. A recent study by Schneider and Borlund (2007) confirms this; in general, different proximity measures are sensitive to small changes in the data and tools must be developed to compare them for a given case study. In fact, much evidence now points to the fact that, in the case of symmetrical (e.g. cocitation) matrices, the normalization of data should be avoided altogether (Leydesdorff & Vaughan, 2006). Nevertheless, a recent improvement to this class of approaches has applied a fitness measure to a *k*-means method, given a dissimilarity matrix, to identify an ideal decomposition scheme (Hsieh & Magee, 2008). It is an interesting technique, since it provides a unique measure of optimality and makes use of the properties of a random network (also used in our approach), but there is still a dependence on the choice of dissimilarity measure used, and the authors restrict themselves to the case unweighted networks.

Recently, two different approaches have largely avoided the use of such "indirect" measures. First, Leydesdorff (2005) has proposed an algorithm to divide up networks using tools from information theory, specifically based on an optimization of the "in-between group uncertainty" (Leydesdorff, 1991). Intuitively, this approach seems reasonable: we would like to find a sequence of divisions that successively minimize the heterogeneity of the grouping scheme. However, we believe that, (1) appeals to information theory are unnecessary when it comes to the topology of networks (the connection between the two is not always clear), and that, (2) based on the results obtained from this approach, the algorithm fails to detect what we would intuitively recognize as community structure (see Section 4, below). Another promising approach in the same vein uses the amount of information involved in a random walk through a weighted network (Rosvall & Bergstrom, 2008). This information flow across the network thus reveals much of its inherent community structure, effectively "coarse-graining" the network so that local communities can be found. We contend that, for networks without explicit "flow", such as undirected cocitation networks, such an approach is not necessary. It is both more useful and more straightforward to examine the topological structure of the network directly. Finally, a different perspective on community structure in various types networks developed by Palla et al. (2005) allows the detection of overlapping communities. The authors show that allowing a node to have different "memberships" allows for greater flexibility and accuracy in several different types of networks, and is thus a net improvement over previous divisive or agglomerative methods. However, the reliance on *k*-cliques and the use of a link weight threshold makes this type of method less suitable to our purposes. Nevertheless, while our hierarchical clustering approach (discussed in the following section) allows one to see the connection between scientific specialties (communities), it would be interesting to have an algorithm which could directly identify nodes (actors) which have a more or less equal attachment to several communities.

## 3. Method

The Girvan-Newman (GN) algorithm (Girvan & Newman, 2002; Newman & Girvan, 2004) is well-known as the canonical method for community detection in complex networks. Essentially, this method consists in cutting links with high values of betweenness (in terms of geodesics passing through a link) and monitoring the graph's modularity $Q$, loosely defined as a measure of how meaningful a given division of the network into subgroups is, while taking into account the number of random links that would be expected within a subgroup. In 2004, Newman expanded this technique to weighted graphs: the basic algorithm remains the same, since a weighted network (multigraph) can be simply mapped onto unweighted graphs (given that we can recover the same adjacency matrix), but with the caveat that the probability of cutting a link is now inversely proportional to its weight. Indeed, it has been recently found that the community structure of networks is highly dependant on weights (Fan et al., 2006, 2007). After a given division into communities, the modularity is computed as,

$$Q = \frac{1}{2m}\sum_{ij}\left(A_{ij} - \frac{k_i k_j}{2m}\right)\delta(c_i, c_j) \quad (1),$$

where $m = (1/2)\sum_{ij} A_{ij}$ is the total number of edges in the graph, $w_{ij}$ is the weight of a given edge between nodes $i$ and $j$, $k_i$ is the degree of node $i$ (the sum of all weights of links involving that node), and $c_i$ is the community assigned to node $i$, such that $\delta = 1$ if $c_i = c_j$ and 0 otherwise (so we are only considering nodes within the same community for the sum). Note that the $\frac{k_i k_j}{2m}$ term represents the expected number of edges in the community given a purely random graph. Clearly, the value of $Q$ goes to zero both when there are no divisions (only one community) and when there are as many nodes as there are communities; it cannot be greater than 1. Modularity is not an appropriate measure for community structures in all networks. In some networks, such as food webs, community structure would be reflected by commonality of sources and targets, not higher proportions of internal links. Cocitation networks, though, are well-suited to the measure. Even a negative citation alongside a positive citation implies some topical similarity between the two papers.

However, a standard implementation of the GN algorithm for weighted cocitation networks is not straightforward and is extremely expensive in computational time. Consider a typical cocitation network (see Fig. 2) where many highly-cited authors are cocited with a plethora of different researchers, including many working outside their community. For instance, a Nobel prize-winning biologist might be cited fairly often with a Nobel prize-winning physicist. Also, certain authors are very central, but are not specific to any community and are thus isolated. Several specific profiles of scientists can explain this behaviour. In our network (Fig. 2), for instance, Johannes Stark (at the bottom-right of the figure) was previously found to have the third highest degree centrality of all physicists during this period (Gingras, 2007), but because of having published in a wide variety of specialties, he finds himself in a small community (with B. Walter).

These types of problems are similar to those faced when applying an algorithm based on the identification of dense subgroups (Schildt & Mattson, 2006). The GN algorithm successfully identifies communities on the periphery of the network, but almost never cuts "heavy" links (high numbers of cocitations), even though this is occasionally necessary in order to bring out the communities inherent in the network. In *real*, relatively compact cocitation networks with "strong" (but divisible) cores where practically everyone is co-cited with everyone at some point in time, an algorithm organized in this way can be of some heuristic value, but will have great difficulty uncovering the optimal community structure.

The algorithm of Blondel et al. (2008) balances optimization of modularity with running time and sensitivity to local structure. Each node is first placed in separate communities. Iterating over all nodes, we check if moving the node from its current community to any community to which a neighbor belongs would yield an increase in modularity. If so, we move the node to the neighboring community that gives the highest increase in modularity and continue the process until equilibrium is reached. Then, we project each community as a single node in a new network, with edges between community-nodes where there were edges between nodes in the communities in the original network. The weights of the new edges are obtained by summing over all previous weights (including self-loops). Finally, the entire process is repeated until there is no change in the community structure. This often occurs when the entire network is part of a single community. The result is a hierarchy of communities for the network. Unlike GN, which requires a hard-to-make decision about where to cut, all levels of the hierarchy reflect a coherent community structure, and only a few levels are generally present, even in very large networks. Our results on cocitation networks only use the very first level of the hierarchy, with the smallest communities, as these tend to reflect the structure of scientific specialties in the networks we have examined.

The algorithm runs quickly, with Blondel et al.'s implementation taking just one hundred and fifty two minutes to run on a network of 118 million nodes and 1 billion edges. Even with this very fast evaluation, the detected community structures tend to have high modularity compared to those detected with other algorithms (Blondel et al., 2008).

## 4. Results and discussion

Following the debate over the use of various similarity measures (Ahlgren et al., 2003, 2004; White, 2003; Leydesdorff & Vaughan, 2006; Leydesdorff, 2008) and other techniques (Leydesdorff, 2005), we begin by testing Blondel et al.'s algorithm on the often-used case of 12 highly-cited authors working in the information retrieval field and 12 from the field of scientometrics or bibliometrics. Fig. 1 displays the original network, along with the results of the community detection algorithm: the three different groups found are differentiated by the shape of their nodes. Group 1 recovers the bibliometrics group in its entirety, while the information retrieval group was divided into two subgroups (Groups 2 and 3). The value of $Q$ obtained from Eq. (1) is quite high, around 0.31. Upon closer inspection of the original cocitation matrix (Ahlgren et al., 2003) or of Fig. 1, we notice that these are the "hard" (Group 2) and "soft" (Group 3) retrievalists (White, 2003). In a Pearson's *r* or cosine analysis, a dendogram also reveals the presence of these two groups of information retrievalists, as well as two less obvious groups of scientometrics researchers. Again, a proximity matrix-based approach would require an *a posteriori* choice as to which subdivisions to include. Therefore, it is particularly interesting that Blondel et al.'s algorithm is able to distinguish between the structures of these two areas.

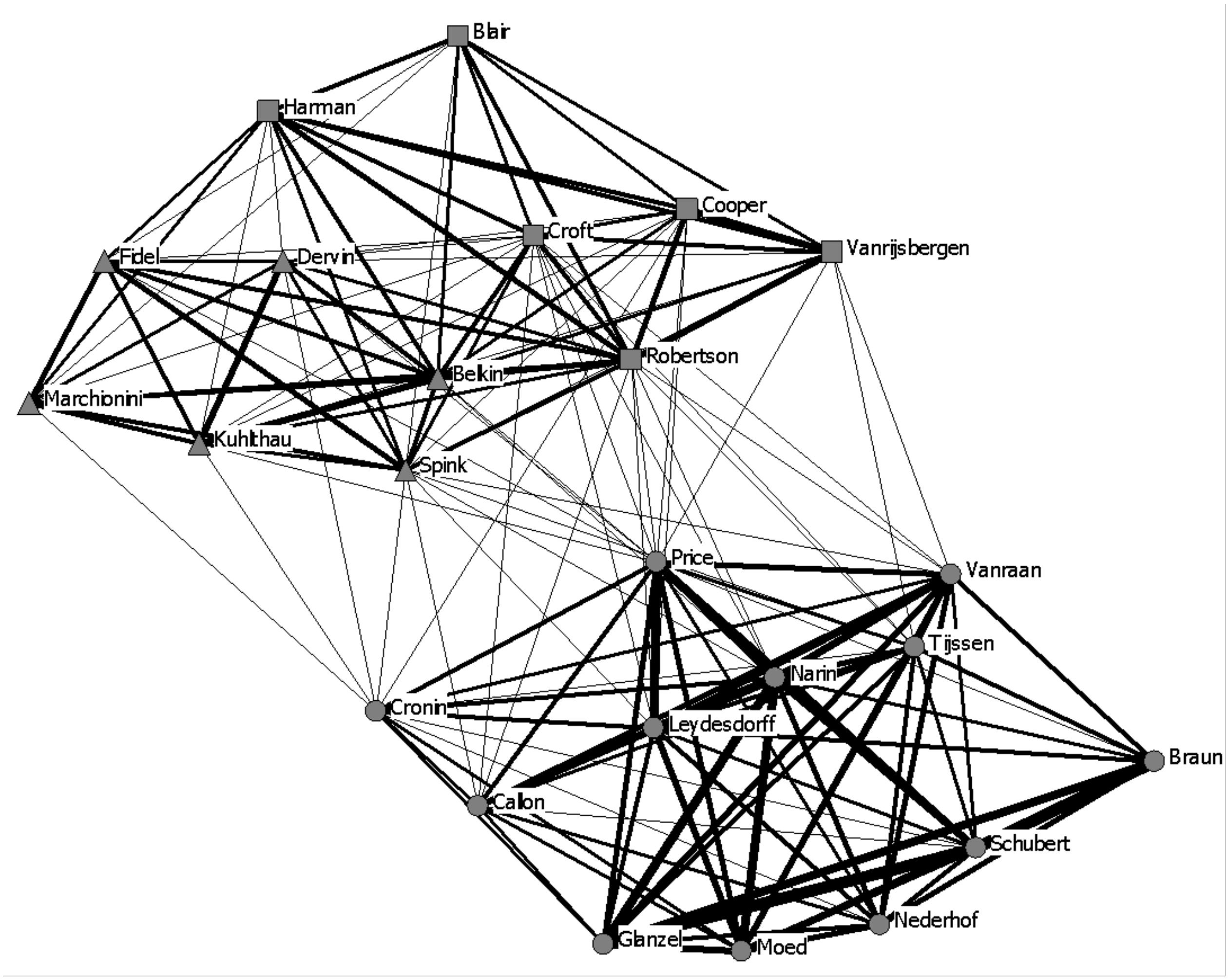

**Figure 1: Network of 24 researchers in the fields of scientometrics and information retrieval. Groups 1, 2 and 3, as found by Blondel et al.'s community detection algorithm, are denoted by circles, squares and triangles, respectively. The links are proportional to the tie strength and the layout is done using a simple repulsion scheme in Netdraw.**

In his information theory-based approach, Leydesdorff (2005) found 13 (much smaller) clusters. Indeed, the author notes that this is counterintuitive, but argues that it more accurately reveals the data's inherent structure. We argue, however, that for a clustering or a community detection algorithm to be useful in cocitation analysis, *the resulting divisions must be conceptually meaningful*. In other words, community detection must provide information on the inherent topology that can be understood in terms of sociology, the history of ideas, science policy, etc. This does not mean that it must correspond to what we already know about the structure – the case of emerging scientific specialties being a prime example (Small, 2006) – only that it must allow for a consistent interpretation. We concur with Leydesdorff on the fact that this particular network – and, to a lesser degree, many other cocitation networks (*cf.* Kreuzmann, 2001) – is problematic in that all the actors have been pre-selected partially according to their "membership". In such cases, a good division into subgroups becomes virtually unavoidable.

In order to perform a more stringent test of the algorithm, we have used networks of much greater complexity, usually composed of the most cited authors in a large subset of science. We have analyzed, for instance, the field of biology (1953-70), physics (1900-1944) (Gingras, 2007), as well as physics, mathematics and chemistry combined (1905-1911 and 1937-1944) and have obtained meaningful and consistent results for each one. Perhaps one of the most "difficult" networks – due to the (presumed) presence of scientific specialties despite a very "close-knit" community as a whole – is that of physics (1905-1911). For visualization purposes, we have restricted our sample to the 100 most cited authors during this period, using the Century of Science database, constructing cocitation links for two or more cocitations. This last restriction has no impact on the results, only on the file sizes and computational time. We have performed an in-house standardization of authors' names (whenever possible) and used only the citing papers from journals which we have defined as pertaining to astrophysics, astronomy, physics and the earth sciences. The raw network is shown in Fig. 2 (grey links and black links combined). In this case, similarity measures do not yield a meaningful MDS picture (the Kruskal stress is very high), nor a very useful dendogram from AHC (Fig. 3), regardless of the specific methods used, or the similarity measure employed, in these two approaches. Though we can recognize many of the same specializations (e.g. relativity, geophysics) from the dendogram, there is no clear way of selecting the "level" at which the groups should be identified. In addition, many of the groups which we would intuitively identify as specialties (see Fig. 2) cannot be clearly identified by AHC or MDS.

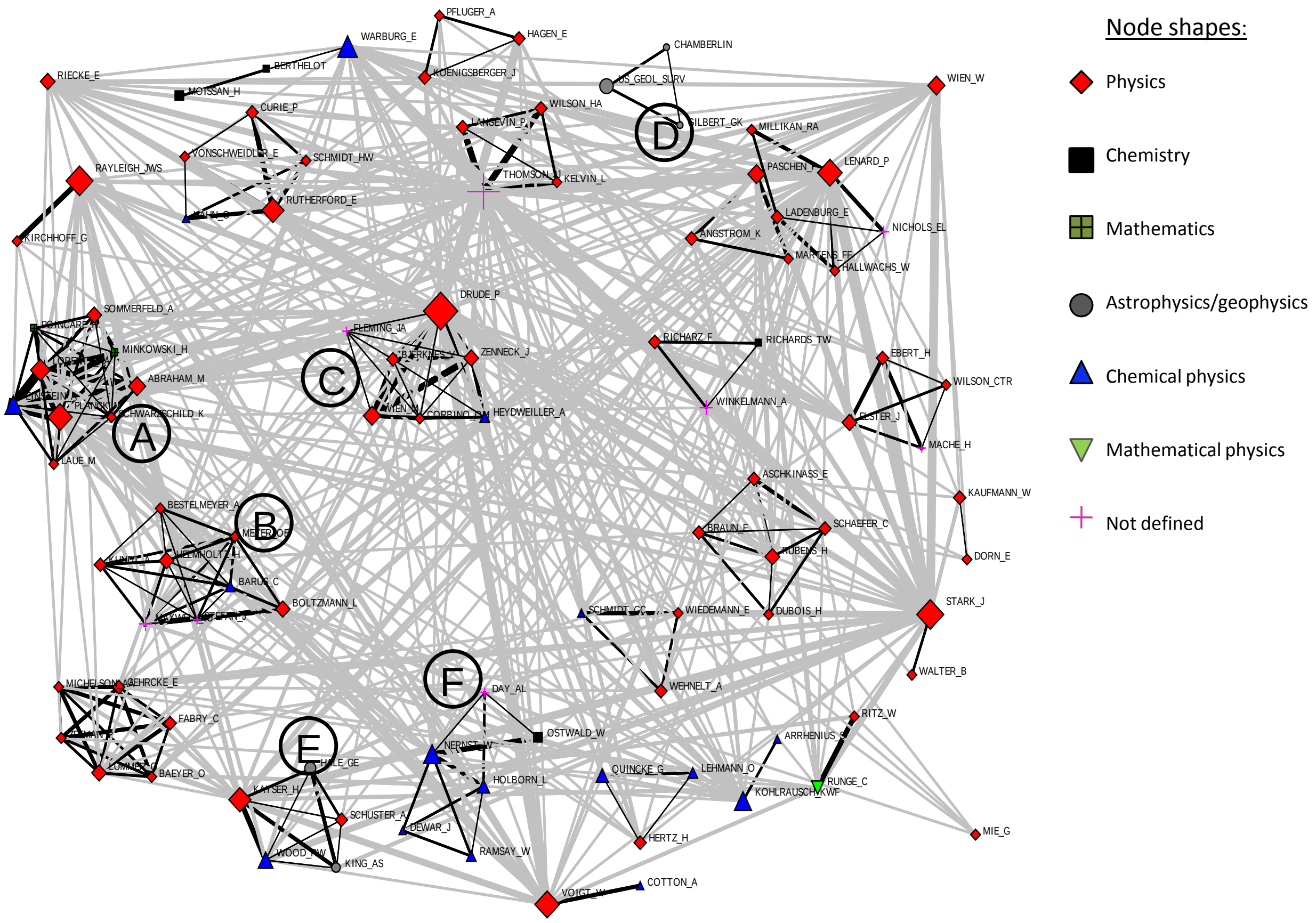

**Figure 2. Complete cocitation network of the top 100 most cited authors in physics journals (1905-1911). The black links are within a given community, while grey links are *between* communities. The width of the ties is proportional to the number of cocitations, and the node size is proportional to the total number of citations received by an author in physics journals during this period. The positions of the communities with respect to each other are arbitrary. A representative sample of communities (A-F) is examined in greater detail in the text.**

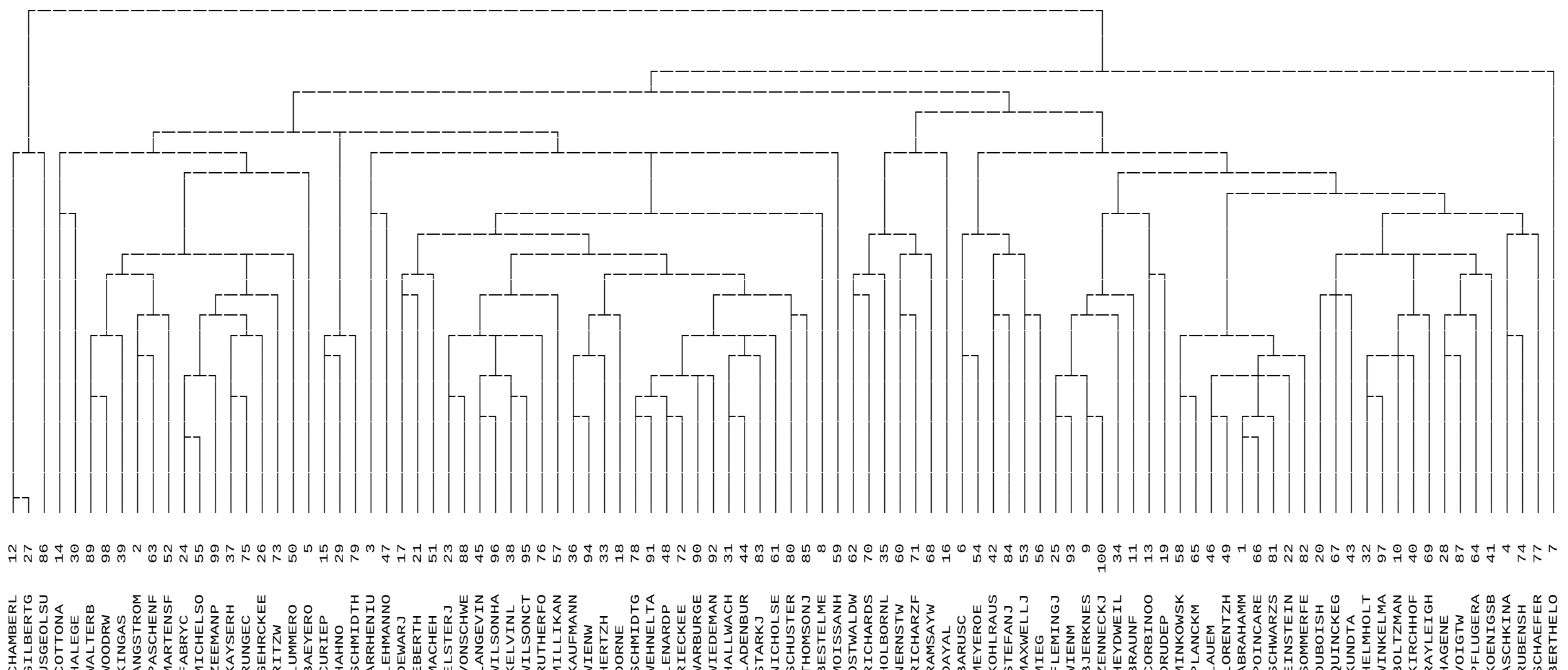

**Figure 3: Dendrogram found using the Pearson's *r* similarity measure. Note that only a few "strong" communities are apparent from visual inspection. The dendograms obtained using the Jaccard or cosine indices are qualitatively similar.**

The results from Blondel et al.'s algorithm, however, reveal that there are readily-detectable communities within this network (Fig. 2). Naturally, the values of *Q* (around 0.023) are much lower than in the previous example, which reflects the number of "heavy" links that are effectively cut by the algorithm. We can confirm, from what we know about the development of physics between 1905 and 1911, that the clusters found do seem to correspond to scientific specialties. We can also check this by examining the frequency of keywords found in the titles of citing articles used to construct the raw network. For instance, the words "relativity" and "radiation" characterize community *A*, "gases" and "pressure" are the most frequent in community *B*, and "electrical" and "oscillations" are found in community *C*. These are indicative of some of the most prevalent topics in physics at this time. We can also verify the algorithm's output and gain insight into the nature of these conceptual communities via their discipline or interdisciplinarity (mathematics, physics, chemistry, astrophysics/geophysics, mathematical physics and chemical physics), by independently looking at which types of journals cite each author most often. In this case, we do not restrict ourselves to physics journals, but use the entire ISI database. Indeed, the various communities found can be loosely described by their disciplinary (or multidisciplinary) membership, even though they are (by definition) all part of physics. Community *D*, for instance, is associated with the earth sciences, community *E* with astrophysics, and community *F* with chemical physics.[1] The last three cases thus find themselves on the periphery of the discipline.

[1] The identification of the discipline of the authors is found from the proportion of his citations in disciplinary journals. For instance, if someone is cited almost exclusively in physics journals, he or she is considered a physicist; if the citations are mostly (and more or less equally) in chemistry and physics journals then he or she is a chemical physicist, etc. This is not an exact characterization, but gives us an idea of the type of specialization, especially for the lesser-known authors.

## 5. Conclusion

We have shown that there are several important advantages in using the algorithm of Blondel et al. (2008) for community detection in cocitation networks. While many other methods discussed can be successfully applied to networks that are small, or where the communities are fairly clear-cut, we believe that a rigorous utilization of cocitation data generally results in much more dense or convoluted networks, and thus requires a more robust approach. Furthermore, we believe that it is imperative that subjective treatment of the data be avoided as much as possible in cocitation analysis. This field of inquiry can greatly benefit from the rapid advances in network analysis, while keeping in mind the primarily conceptual (and not social, for instance) nature of cocitation networks at all times. It would also be interesting to see what other new techniques, such as that of Rosvall and Bergstrom (2008), can reveal about the intellectual structure of a scientific domain.

We have established that the resulting communities detected from our network consistently correspond to what we would intuitively recognize as scientific specialties or invisible colleges (if we restrict ourselves to the active researchers in the cocitation network). Our algorithm unambiguously reveals that, in terms of the hierarchy of different community sizes, specialties are the smallest subset of authors that can be "brought together". These techniques could be of great use to historians or sociologists of science, by tracking the emergence, demise, proximity or fusion of specializations, as well as the evolution of scientific paradigms (Mullins, 1972; Chubin, 1976; Mullins et al, 1977; Small, 2006). Given a specific community, we can identify – using keywords, for instance – its ideas, methods and membership. Most importantly, this is done from data that is not "tailored" in advance to ensure that these subgroups are visible.

## Acknowledgements

We thank Vincent Larivière and Katy Börner for useful discussions and insightful comments during the preparation of this manuscript.

## References

Ahlgren, P., Jarneving, B. & Rousseau, R. (2003). Requirement for a cocitation similarity measure, with special reference to Pearson's correlation coefficient. Journal of the American Society for Information Science and Technology, 54(6), 550-560.

Ahlgren, P., Jarneving, B. & Rousseau, R. (2004). Author cocitation analysis and Pearson's r. Journal of the American Society for Information Science and Technology, 55(9), 843.

Bayer, A.E., Smart, J.C. & McLaughlin, G.W. (1990). Mapping intellectual structure of a scientific subfield through author cocitations. Journal of the American Society for Information Science, 41(6), 444-452.

Bensman, S.J. (2004). Pearson's r and author cocitation analysis: A commentary on the controversy. Journal of the American Society for Information Science and Technology, 55(10), 935-936.

Blondel, V., Guillaume, J.-L., Lambiotte, R. & Lefebvre, E. (2008). Fast unfolding of community hierarchies in large networks. arXiv:cond-mat/0803.0476

Boyack, K.W., Börner, K.& Klavans, R. (2007). Mapping the structure and evolution of chemistry research. Proceedings of the 11th International Conference of the International Society for Scientometrics and Informetrics, 112-123.

Chen, C. (2004). Searching for intellectual turning points: Progressive knowledge domain visualization. Proceedings of the National Academy of Science, 101(S1), 5303-5310.

Chubin, D.E. (1976). The conceptualization of scientific specialties. Sociological Quarterly, 17(4), 448-476.

Crane, D. (1972). Invisible Colleges. Chicago, University of Chicago Press.

Fan, Y., Li, M.H., Zhang, P., Wu, J.S. & Di, Z.R. (2007). The effect of weight on community structure of networks. Physica A, 378, 583-590.

Fan, Y., Li, M.H., Zhang, P., Wu, J.S. & Di, Z.R. (2006). Accuracy and precision of methods for community identification in weighted networks. arXiv:cond-mat/0607271.

Gingras, Y. (2007). Mapping the changing centrality of physicists (1900-1944). Proceedings of the 11th International Conference of the International Society for Scientometrics and Informetrics, 314-320.

Girvan, M. & Newman, M.E.J. (2002). Community structure in social and biological networks, Proceeding of the National Academy of Sciences of the United States of America, 99, 7821-7826,

Gmür, M. (2003). Co-citation analysis and the search for invisible colleges: A methodological evaluation. Scientometrics, 57(1), 27-57.

Hsieh, M.-H. & Magee, C.L. (2008). An algorithm and metric for network decomposition from similarity matrices: Application to positional analysis. Social Networks, 30, 146-158.

Klavans, R. & Boyack, K.W. (2006). Quantitative evaluation of large maps of science. Scientometrics, 68(3), 475-499.

Kreuzman, H. (2001). A co-citation analysis of representative authors in philosphy: Examining the relationship between epistemologists and philosphers of science. Scientometrics, 51(3), 525-539.

Leydesdorff, L. (1991). The static and dynamic analysis of network data using information theory. Social Networks, 13, 301-345.

Leydesdorff, L. (2005). Similarity measures, author cocitation analysis and information theory. Journal of the American Society for Information Science and Technology, 56(7), 769-772.

Leydesdorff, L. & Vaughan, L. (2006). Co-occurrence matrices and their applications in information science: Extending ACA to the Web environment. Journal of the American Society for Information Science and Technology, 57(12), 1616-1628.

Leydesdorff, L. (2008). On the normalization and visualization of author co-citation data: Salton's cosine *versus* the Jaccard index. Journal of the American Society for Information Science and Technology, 59(1), 77-85.

Marshakova, I.V. (1973). A system of document connections based on references. Scientific and Technical Information Serial of VINITI, 6, 3-8.

Mullins, N.C. (1972). The development of a scientific specialty: the phage group and the origins of molecular biology. Minerva, 10, 52-82

Mullins, N.C., Hargens, L.L., Hecht, P.K. & Kick, E.L. (1977). The group structure of cocitation clusters: A comparative study. American Sociological Review, 42(4), 552-562.

Newman, M.E.J. (2004). Analysis of weighted networks. Physical Review E, 70, 056131.

Newman, M.E.J. & Girvan, M. (2004), Finding and evaluating community structure in networks. Physical Review E, 69, 026113.

Palla, G., Imre, Derényi, I., Farkas, I. & Viscek, T. (2005). Uncovering the overlapping community structure of complex networks in nature and society. Nature, 435, 814-818.

Rosvall, M. & Bergstrom, C.T. (2008). Maps of random walks on complex networks reveal community structure. Proceedings of the National Academy of Science, 105(4), 1118-1123.

Schaeffer, S.U. (2007). Graph clustering. Computer Science Review, 1(1), 27-64.

Schildt, H.A. & Mattsson, J.T. (2006). A dense network sub-grouping algorithm for co-citation analysis and its implementation in the software tool *Sitkis*. Scientometrics, 67(1), 143-163.

Schneider, J.W. & Borlund, P. (2007). Matrix comparison, Part 1: Motivation and important issues for measuring the resemblance between proximity measures or ordination results. Journal of the American Society for Information Science and Technology, 58(11), 1586-1595.

Small, H. (1973). Co-citation in the scientific literature: A new measure of the relationship between two documents. Journal of the American Society for Information Science and Technology, 24, 265-269.

Small, H. (1978). Cited documents as concept symbols. Social studies of science, 8(3), 327-340.

Small, H. (2006). Tracking and predicting growth areas in science. Scientometrics, 68(3), 595-610.

Small, H. & Griffith, B.C. (1974). The structure of scientific literatures, I. Identifying and graphing specialties, Science Studies, 7, 17-40.

Small, H. & Sweeney, E. (1985). Clustering the *Science Citation Index*® using co-citations: 1. A comparison of methods.

White, H.D. (1981). Cocited author retrieval online: An experiment with the social indicators literature. Journal of the American Society for Information Science, 32, 16-22.

White, H.D. (2003). Author cocitation analysis and Pearson's r. Journal of the American Society for Information Science and Technology, 54(13), 1250-1259.

White, H.D. & McCain, K.W. (1981). Author co-citation: a literature measure of intellectual structure. Journal of the American Society for Information Science, 32, 163-172.